\documentclass[aps,prl,amssymb,amsmath,twocolumn,floatfix]{revtex4}
\usepackage{graphicx}
\usepackage{amsmath}

\begin{document}
\title{Dynamic fractals in spatial evolutionary games}
\author{Sergei Kolotev$^{1, 2}$}
\author{Aleksandr Malyutin$^{1,2}$}
\author{Evgeni Burovski$^{1, 2}$}
\email{evgeny.burovskiy@gmail.com}

\author{Sergei Krashakov$^{2, 3}$}
\author{Lev Shchur$^{1, 2, 3}$}
\affiliation{$^1$ National Research University Higher School of Economics, 101000 Moscow, Russia}
\affiliation{$^2$ Science Center in Chernogolovka, 142432 Chernogolovka, Russia}
\affiliation{$^3$ Landau Institute for Theoretical Physics, 142432 Chernogolovka, Russia}

\begin{abstract}
We investigate critical properties of a spatial evolutionary game based on the Prisoner's Dilemma.
Simulations demonstrate a jump in the component densities accompanied by drastic changes in average sizes of the component clusters.
We argue that the cluster boundary is a random fractal. Our simulations are consistent with the fractal dimension of the boundary being equal to 2, and the cluster boundaries are hence asymptotically space filling as the system size increases.
\end{abstract}

\maketitle

The surge of interest in game theory can be traced to the
seminal works of John Nash in the middle of the 20th century. The main
subject of classical game theory is finding the optimal strategy in games
between two or more individuals (players or agents), where each individual has several
possible behaviors. A repeated game is a situation where the same agents play the game with the
same rules multiple times. Rational behavior of an agent then evolves
with time based on the memory of past encounters. An agent's strategy thus evolves, the so-called evolution of cooperation \cite{Axelrod81, Axelrod06}.

A prototypical model in game theory is the so-called Prisoner's Dilemma,
played by two agents in discrete time steps. In each round of the game, each
agent uses one of two possible strategies, \emph{cooperate} $\mathcal{C}$ or
\emph{defect} $\mathcal{D}$, and receives a payoff that depends
on the strategies of the agent and its opponent~\cite{Tadelis2013}.

Evolutionary game theory (see, e.g., \cite{Jonker78, Smith82,
Weibull95, Nowak06} and the references therein) investigates the behavior of
large populations, where a macroscopic number of agents use
a finite number of strategies. While classical game theory deals with
individual agents, evolutionary game theory focuses on the winning strategies
themselves rather than individuals. Spatial evolutionary games are played with agents
arranged in some spatial structures and interacting with other agents
in their immediate neighborhoods. Various geometries have been explored,
including regular grids \cite{Nowak92, Nowak93}, random graphs and small world
networks \cite{Szabo05}, and evolving random graphs \cite{Perc09, Helbing2013}.

The spatial arrangement of agents yields emergent geometric structures---groups of
agents who synchronize their behaviors with their neighbors and
compete with other groups. The temporal evolution of these geometric structures can
be highly nontrivial.

In this letter, we study a simple version of an evolutionary game
based on the Prisoner's Dilemma \cite{Nowak92, Nowak93}. The game is
deterministic, and the time evolution is governed by a single parameter, the
payoff ratio. Although the local rules are apparently simple, the steady state of
the game features a series of very different dynamic regimes separated by sharp
transitions. We characterize the geometric properties of the emergent structures
across transitions.

We obtained several results that might be surprising for the statistical physics community.
We found that transitions between steady states of the structures
are not related to any kind of transitions known in statistical mechanics. The transitions are sharp
but are not similar to first-order thermodynamic transitions~\cite{Binder-review}. Clusters of agents with similar strategies do percolate
from boundary to boundary of the finite systems investigated. And in contrast to the percolation clusters in thermodynamic equilibrium~\cite{Review-universality, Thouless}, the dimension of the fractals is 2 in the plane. A cluster boundary looks very irregular, and the dimension of the boundary is again equal to the dimension of the space.

Following Refs.\ \cite{Nowak92, Nowak93}, we define the game rules as follows:
$L^2$ agents are arranged on an $L\times L$ rectangular grid in two dimensions. The game is globally synchronous and
is played in discrete time steps. At each time step,
an agent interacts with its eight neighbors (the chess king's moves) and
itself
\footnote{The presence of self-interaction can be motivated by considering an
agent to represent a group of individuals. Self-interaction then
encapsulates some internal dynamics of this group. The qualitative features of the
model are independent of the inclusion of self-interaction.}.
The total score of an agent in a round is the sum of the payoffs of all nine
games played in the current round.
When all pairwise games are played and all payoffs are known, agents change
their strategies for the next round. Various adaptation behaviors are possible.
We use the simplest case of maximally opportunistic agents with a short memory: at each time
step, an agent adopts the strategy with the
maximum payoff among itself and its opponents in the preceding round. For two agents, this strategy
is trivial. It becomes more interesting when the maximally opportunistic
adaptation is used in a spatial evolutionary context.

The payoff an agent receives in an elementary game depends on the strategies
of the agent and its opponent. We use the following payoff structure
\cite{Nowak92, Nowak93}: (i) If both agents defect, they receive nothing.
(ii) If both agents cooperate, each of them receives a payoff of $S$, which
we set to $S=1$ without loss of generality. (iii) In the interaction of
$\mathcal{C}$ and $\mathcal{D}$, the defector receives a payoff $T>S$ and the
cooperator receives zero. The payoff structure hence depends on only one
parameter, the payoff ratio $b=T/S$.

The spatial game can obviously be described as a cellular automaton with a
particular set of transition rules, but the description in terms of
cellular automata turns out to be very complex: the state of an agent
in the next round depends on the payoffs of its neighbors, which in turn depend
on their neighbors. Because 25 agents are relevant, the transition
table size is $2^{25}$, in contrast to the transition matrix
for Conway's Game of Life, which has $2^9=512$ rules.

\emph{Qualitative analysis.---} The game is deterministic, and the full time
evolution is completely defined by the initial conditions (the spatial
distribution of strategies $\mathcal{C}$ and $\mathcal{D}$ at $t=0$) and
the value of the payoff parameter $b$.
The discrete structure of the payoffs leads to a series of very different
dynamic regimes separated by sharp transitions at special values of $b$.
Moreover, for fixed initial conditions, the dynamics is exactly identical for all
values of $b$ between these transition points. Statistical fluctuations enter via
our use of random, unstructured initial conditions: physical observables are
calculated as averages over both time evolution in the steady state (which is
deterministic given initial conditions) and the ensemble average over a set of
realizations of initial conditions (where the steady states are equivalent
in the statistical sense).

It is instructive to consider the time evolution of
small local objects, i.e., clusters of one strategy embedded into a sea of the
other strategy. For $b<1$, defectors always lose. For $b>3$, cooperators unconditionally win.
For $1<b<9/5$, a zoo of various small objects (gliders, rotators, etc.)\ are
possible. Small clusters of $\mathcal{D}$ remain small and large
clusters of $\mathcal{D}$ shrink. Conversely, for $b>9/5$, a 2$\times$2 or larger
cluster of $\mathcal{D}$ grows.
The situation is reversed for defectors: a 2$\times$2 cluster of $\mathcal{C}$ grows
for $b<2$, while a large cluster of $\mathcal{C}$ shrinks for $b>2$.

Therefore, $9/5<b<2$ is the fierce competition regime where clusters of
$\mathcal{C}$ can grow in regions of $\mathcal{D}$ and vice versa. Starting
from a single defector in a center of a large game field, the steady state is
a dynamic fractal with a well-defined average density of $\mathcal{D}$,
the ``evolutionary kaleidoscope" in Ref.\ \cite{Nowak93}.

\emph{Numerical simulations.---} To study the time evolution of the game with
random, unstructured initial conditions, we use a direct numerical simulation.
We arrange $L^2$ agents on an $L\times L$ square grid with $L$ up to 1000. We
use periodic boundary conditions to minimize edge effects. We use up to
2$\times$10$^5$ time steps for small lattices and up to 2$\times$10$^3$ time steps
for larger lattices. For the initial state of the game field at $t=0$, we assign
the strategy $\mathcal{C}$ to an agent with a fixed probability $p_i$.
We consider up to 100 realizations of the initial conditions (replicas).
To compute steady-state averages, we discard the first 10$^3$ iterations
for equilibration.

\begin{figure*}
\includegraphics[width=0.33\textwidth, keepaspectratio=True]{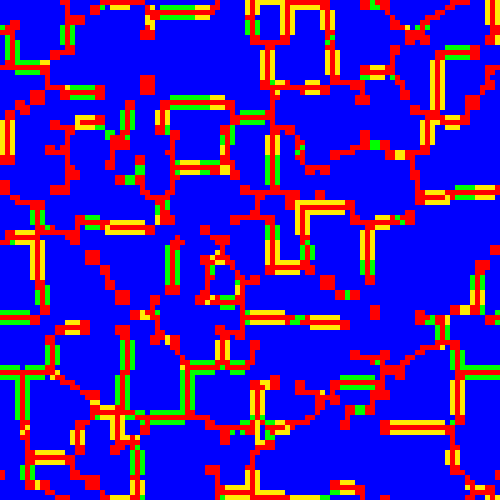}~%
\includegraphics[width=0.33\textwidth, keepaspectratio=True]{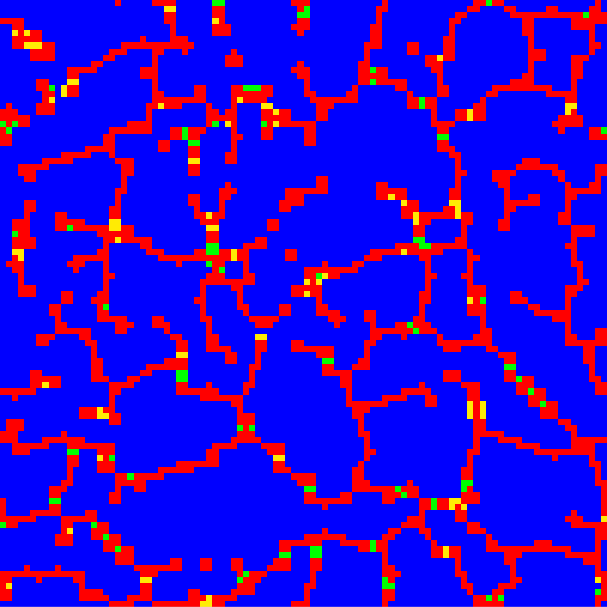}~%
\includegraphics[width=0.33\textwidth, keepaspectratio=True]{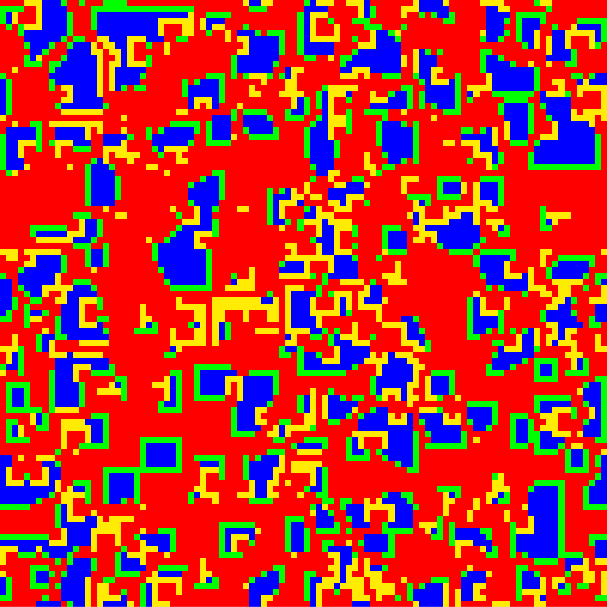}
\caption{(Color online.) Representative snapshots of the game field for
$b=1.74$ (left), $b=1.79$ (center) and $b=1.81$ (right). The color coding is
consistent with Ref.\ \cite{Nowak92, Nowak93}: blue is $\mathcal{C}$, red is
$\mathcal{D}$, yellow is a $\mathcal{D}$ that was
a $\mathcal{C}$ in the preceding round, and green is a $\mathcal{C}$ that was
a $\mathcal{D}$ in the preceding round. See text for discussion.
}
\label{fig:snapshots}
\end{figure*}

In agreement with Refs.\ \cite{Nowak92, Nowak93}, we find that typical
configurations of the game field change drastically across the critical value
of $b_c=9/5$. For $b<b_c$, cooperators form relatively static web-like
structures spanning the entire game field; for $b>b_c$, the game field
features ``blobs" of various sizes (see Fig.\ \ref{fig:snapshots}).
Clusters of
both $\mathcal{C}$ and $\mathcal{D}$ move, grow, and collide chaotically, leading
to the game field changing at time scales of the order of several time
steps.

\begin{figure}
\includegraphics[width=\columnwidth, keepaspectratio=True]{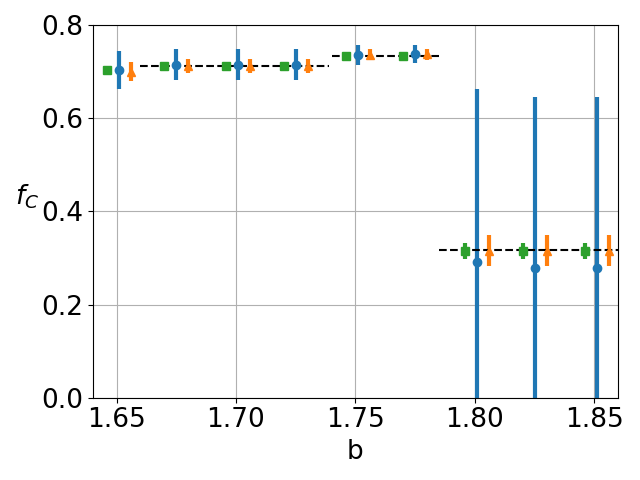}
\caption{Density of cooperators $f_C$ as function of the payoff $b$ with lattice
sizes 20$\times$20 (circles), 50$\times$50 (triangles), and 100$\times$100 (squares).
All simulations are performed at $b = 1.651, 1.675, 1.701, 1.725, 1.751, 1.775,
1.801, 1.825, 1.851$ (for clarity,
triangles are shifted horizontally slightly to the right and squares, to the left).
Error bars are shown for all points and reflect averaging over 25 independent realizations of
the initial conditions, each simulated for 2$\times 10^4$ generations.
Dashed lines are to guide the eye. We note that for $b>9/5$, the average density agrees with
the magic value $f_C=12\,\log{2}-8\approx0.318$ \cite{Nowak93}.
See the text for discussion.}
\label{fig:coop_dens}
\end{figure}

To quantify the apparent changes of the game dynamics as $b$ varies,
we compute the average density of cooperators in the steady state for a range of
payoffs $b>1$. For each value of $b$, we take 25 independent realizations of
the initial conditions and compute time averages of the density of cooperators
discarding the first $10^3$ time steps for equilibration and averaging over
up to $2\times 10^4$ steps. Figure~\ref{fig:coop_dens} shows the
results of these simulations. We clearly see a sharp transition around the
predicted value $b_c=9/5$, where the average steady state density of
$\mathcal{C}$ drops from $\sim 0.7$--$0.9$ to $\sim 0.3$.

Several features stand out in Fig.\ \ref{fig:coop_dens}. First, not only
the average value changes at $b=b_c$, but also the spread of individual
measurements increases dramatically. The spread itself
is clearly a finite-size effect, which progressively decreases as
$L$ increases. Second, the average density $f_C$ for $b>b_c$ agrees well with
the value $f_C = 12\log{2}-8\approx0.32$, found in \cite{Nowak93} for regular
``evolutionary kaleidoscopes," which develop from an initial state with a single
cooperator and $L^2-1$ defectors.

\begin{figure}
\includegraphics[width=\columnwidth, keepaspectratio=True]{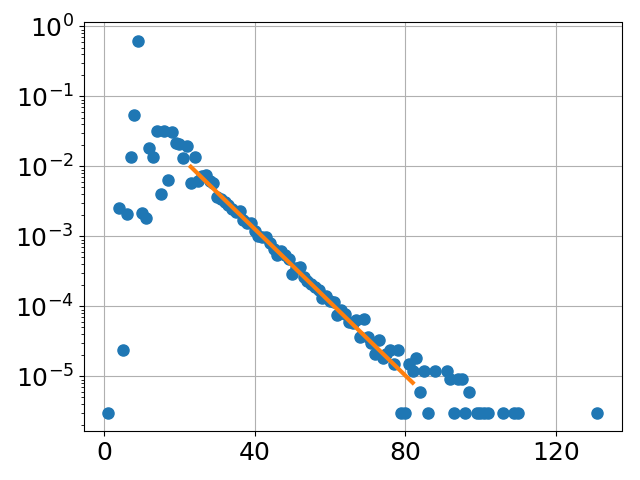}
\includegraphics[width=\columnwidth, keepaspectratio=True]{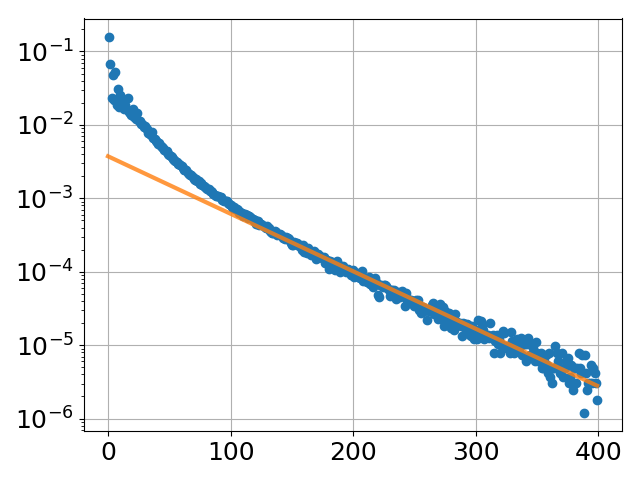}
\caption{Cluster mass distributions $w(m)$ normalized such that
$\sum_m w(m)=1$, for $b=1.79$ (top) and $b = 1.81$ (bottom). Here, $L = 100$, and
the initial condition is that an agent is a defector with probability $p_i=0.21$.
The simulations are done for 100 independent realizations of initial conditions.
Each realization is simulated for 100 (1000) steps for $b=1.79$ ($b=1.81$) after
a 1000 steps for equilibration. Longer simulation times do not change the picture,
and the results are independent of the system size.
See the text for discussion.}
\label{fig:cluster_size_distribution}
\end{figure}

\emph{Cluster size distribution.---} To characterize the transition at $b=b_c$,
we perform the following simulation. At each time step, we decompose the game
field into connected clusters of $\mathcal{C}$ and $\mathcal{D}$, and record
the ``mass" (i.e., the number of sites) of each cluster.
Figure\ \ref{fig:cluster_size_distribution} shows the distribution $w(m)$
of masses of clusters of $\mathcal{D}$ for $b<b_c$ and $b>b_c$, collected over
1000 time steps of 100 independent realizations of initial conditions for $L=100$.
For $b<b_c$, clusters larger than the system size are virtually nonexistent,
and the distribution has a maximum at the size of around 10 sites. The middle
part of the distribution, for $20<m<80$, is well fit by an exponential function
$w(m)\sim B\exp(-\lambda m)$ with the best-fit values $B = 0.16(2)$ and
$\lambda = 0.12(1)$ (numbers in parentheses represent the fit error bars in units
of the last digit). We now stress that the smoothness of the distribution is a
result of averaging over the initial conditions: the cluster size distribution
for each particular replica is noisy and does not display any discernible structure.

For $b>b_c$, the situation is markedly different: the distribution is monotonic,
where the initial faster-than-exponential drop for $m\lesssim L$
is followed by a long tail that is well fit by an exponential decay
$w(m>100)\sim B\exp(-\lambda m)$ with
$B=3.7(2)\times10^{-3}$ and $\lambda=0.018(2)$.
Here, the time averaging is much more effective than for $b<b_c$ because the game
field changes substantially at the time scale of several time steps.

\textit{Cluster boundaries.---} We again decompose the game field into connected
clusters at each time step. We then record the total length of the interface between
the areas occupied by $\mathcal{C}$ and $\mathcal{D}$. We define the interface
length $p$ as the number of bonds connecting agents of different kinds
(the total number of bonds is $2L^2$ on an $L\times L$ game field with periodic boundary conditions).
This definition of the interface length clearly depends on the system size. A natural
expectation is that $p$ scales as some power $\theta$ of the system size $L$. Several
scenarios are possible. If we naively regard clusters of strategies as droplets of
immiscible liquids, then we expect $p\propto L$. A space-filling curve
would have $p\propto L^2$. A power-law scaling with a noninteger exponent
would indicate that the interface is fractal~\cite{Wang2017}.

For each value of $b$, we simulate for a range of $L$ and measure
the steady-state average value of $p$. We then fit the results with a power law
$p(L)\sim A\times L^\theta+c$ with $A$, $c$, and $\theta$ as fitting parameters.
Here, $A$ is the amplitude, and we also include the free term $c$.
We expect the scaling exponent $\theta$ to differentiate between the regimes $b<b_c$
and $b>b_c$. The results are summarized in Table\ \ref{table:perimeter}. We
find that the amplitude $A$ depends on $b$ only weakly. The value of the parameter $c$ somehow
reflects what can be seen in Fig.\ \ref{fig:snapshots}, i.e., that the effective correlations
are of the order of the lattice spacing in the left figures and larger in the right figure. Most surprisingly,
the scaling exponent $\theta$ is consistent with $\theta=2$ \emph{for all values} of $b$.

\begin{table}
\begin{tabular}{rrrr}
$b$ & $\theta$ & $A$ & $c$ \\
\hline
1.81 & 1.99(1) & 0.35(1) & -20(05) \\
1.79 & 1.99(1) & 0.34(1) & -13(11) \\
1.74 & 2.07(5) & 0.22(4) & 5(9) \\
1.64 & 2.03(2) & 0.29(2) & -5(6) \\
1.49 & 2.00(3) & 0.30(3) & -2(8) \\
1.39 & 2.02(3) & 0.24(3) & 2(6) \\
1.32 & 2.03(2) & 0.18(2) & 5(4) \\
1.28 & 1.96(5) & 0.26(5) & -4(9) \\
1.19 & 1.98(4) & 0.24(4) & -1(7) \\
\end{tabular}
\caption{Interface length as a function of $L$. Fit errors are shown
in units of the last digit. For each value of $b$, we simulate with
$L$ from $10$ to $200$ and fit the results with $p(L)\sim A\times L^\theta+c$.
See the text for discussion.}
\label{table:perimeter}
\end{table}

To double-check this result, we further calculate using a
more traditional definition of the fractal dimension of the interface. Namely,
we use the standard definition of the fractal dimension of a closed set in two
dimensions, the so-called Minkowski dimension~\cite{BisoiMishra2001,JiangLiu2012}.
Let $N(\ell)$ be the minimum number of boxes necessary to completely cover
the cluster interface with boxes of linear extent $\ell$. Then the Minkowski dimension is defined as
\begin{equation}
d_s=\lim_{\ell\to0}\frac{\log N(\ell)}{-\log\ell}.
\label{minkowski_dim}
\end{equation}

Because the game field is inherently discrete, we use the following procedure. We cover the
interface with boxes of increasing linear size $\ell$ and linearly fit the logarithm
of the number of covering boxes as a function of logarithm of $\ell$. In our fitting
procedure, we discard both the smallest box sizes of the order of several lattice spacings
(because the discreteness of the lattice is essential at these scales) and
the largest box sizes of the order of $L$ (because any curve is space filling at these length scales).

\begin{figure}[!h]
\includegraphics[width=\columnwidth, keepaspectratio=True]{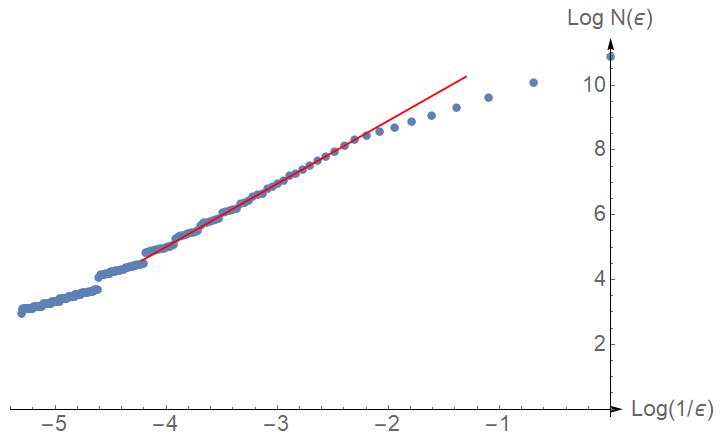}
\caption{Box counting for an interface with $b=1.81$ and $L=200$.
Other system sizes and payoff parameters are similar. Points are the number
of square boxes of a given linear extent needed to completely cover the interface.
The straight line is a linear fit $\log N(\ell)\propto1.95\,\log{(1/\ell)}$
in the window $-4.2<\log(1/\ell)<-2.6$. See the text for discussion.}
\label{fig:minkowski181}
\end{figure}

We simulate 10 independent realizations of the initial configuration with
the probability $p_i=0.21$ of an agent being a defector. For each run,
we take 20 snapshots, separated by 10 time steps. For each snapshot,
we use \eqref{minkowski_dim} to extract the Minkowski dimension.
The fitting procedure is illustrated in Fig.~\ref{fig:minkowski181}
for the system size $L=200$ and the parameter value $b=1.81$.
The results averaged over both time and initial conditions are reported in Table\ \ref{table:minkowski}.

%We choose appropriate window of the box sizes in which results of the measurements can be approximated with the straight line, in this case $-4.2< \log(1/\epsilon)< -2.6$ which is enough for the estimation of $d_s$.

\begin{table}
\begin{tabular}{rrrr}
$L$ & $d_s (b = 1.79)$ & $d_s (b=1.81)$ \\
\hline
100 & 1.776(1) & 1.380(1) \\
200 & 1.762(1) & 1.762(1) \\
500 & 1.936(1) & 1.936(1) \\
1000 & 1.957(1) & 1.957(1) \\
\end{tabular}
\caption{Minkowski dimension of the interface. The numbers in parentheses
show the errors in units of the last digit and include both fitting errors
and statistical variations between measurements.}
\label{table:minkowski}
\end{table}

We find that the values of both the scaling exponent of the cluster interface $\theta$
(see Table~\ref{table:perimeter}) and the Minkowski dimension of the cluster interface $d_s$
tend to the limit value of 2 as $L\to\infty$. In other words, the cluster boundary
(and the cluster interface) are not lines but rather scale as the surface area.
In the pictures at the center and the right in Fig.~\ref{fig:snapshots},
it is not easy to see that all three different objects (blue regions, red regions,
and the boundary between them) have the properties of surfaces.
Nevertheless, the analysis presented here supports this nonobvious fact.

The simulated system in the range of parameters investigated always reaches
some state (after a sufficient relaxation from the initial state),
which is either steady-state or almost steady-state with the small local details
at the boundaries being cyclic with a few time steps and not influencing
the discussed global geometry. The result of the analysis is that
the geometric structures emerge as the steady (or almost steady) state
of the complex dynamic process. The rules are local, but the steady-state
structures demonstrate some global behavior.

There are examples of regular fractals with boundaries described
by the fractal dimension 2: Julia sets and the boundary of the
Mandelbrot set~\cite{Shishikura1998}. In our case, the steady-state
structures are quite random self-organized structures, not regular fractals.
At the same time, to the best of our knowledge, there have been no previous
examples of random self-organized structures with the cluster interface
filling the space as we have shown here.

We stress that our structures shown in Fig.~\ref{fig:snapshots} look similar
to those emerging in various examples of percolation, including discontinuous
percolation (see, e.g., the short review~\cite{HHerrmann2015}) and the
mixing-phase transition~\cite{Bar-Mukamel}, but the interfaces in those examples
never scale with the exponent 2. Another difference is that the behavior changes
only for a discrete set of the control parameter value $b$ and we do not have
a distance to these critical values, as with usual critical phenomena~\cite{Review-universality}.

\begin{acknowledgments}
We thank Lev Barash for the discussion on the fractal dimension of regular fractals.
This work was supported by the Russian Foundation for Basic Research
(Grant No.~16-07-01122, development of algorithms for simulations)
and the Russian Science Foundation (Grant  No.~14-21-00158).
\end{acknowledgments}

%%%%%%%%%%%%%%%%%%%%%%%%%%%%%%%%%%%%%%%%%%%%%%%%%%%%%%%%%%%%%%%%%%%%%%%%%%%%%

\end{document}